\documentclass[11pt]{article}

\usepackage[latin1]{inputenc}
\usepackage[T1]{fontenc}
\usepackage[english]{babel}

\usepackage{hyperref}
\newcommand{\clickhere}{{\raisebox{-.5em}[0em]{\!\includegraphics[height=1.2em]{click_to_play_mouse.pdf}}}}
\newcommand{\click}[1]{\href{http://www.cmm.uchile.cl/~schabanel/2DMINORITY/#1}{\clickhere}}

\usepackage{geometry}                
\usepackage[parfill]{parskip}    
\usepackage{fullpage}
\usepackage{graphicx}
\usepackage{amssymb,amsmath}
\usepackage{epstopdf}
\usepackage{color}
\newcommand{\Minority}{\textbf{Minority}\/}
\newcommand{\OT}{Outer-Totalistic~\textbf{976}\/}

\newcommand{\expect}{{\ensuremath{\operatorname{\mathbb E}}}}

\renewcommand{\leq}{\leqslant}
\renewcommand{\geq}{\geqslant}

\renewcommand{\mod}{\operatorname{\,mod}}

\newcommand{\opmath}[1]{{\ensuremath{\operatorname{#1}}}}
\newcommand{\XOR}{\oplus}
\newcommand{\hull}{{\opmath{\mathsf{hull}}}}
\newcommand{\dmin}{{\opmath{\delta}}}
\newcommand{\dout}{{\opmath{\hat\delta}}}
\newcommand{\dhull}{{\opmath{\bar\delta}}}
\newcommand{\checkbb}{{\raisebox{-.1em}{\includegraphics[height=.8em]{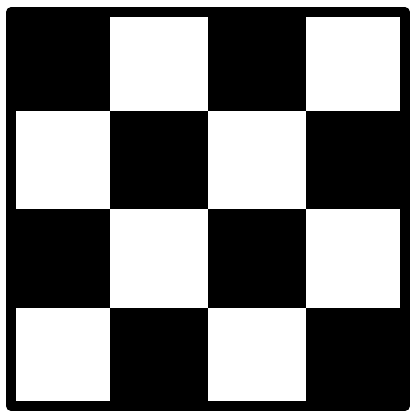}}}}
\newcommand{\checkb}{{\raisebox{-.1em}{\includegraphics[height=.8em]{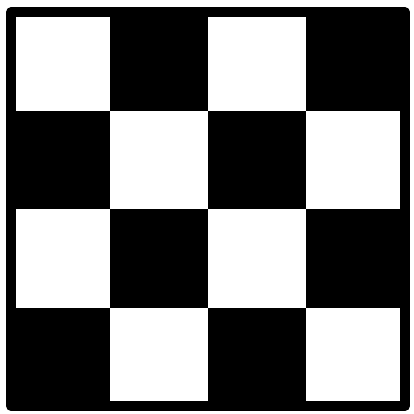}}}}
\newcommand{\fatcheckb}{{\raisebox{-0.05em}{\includegraphics[height=.6em]{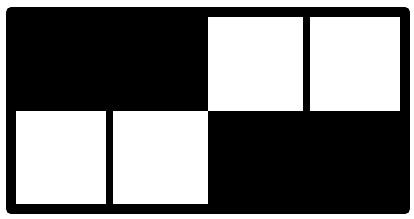}}}}
\newcommand{\linecheckb}{{\raisebox{-0.05em}{\includegraphics[height=.6em]{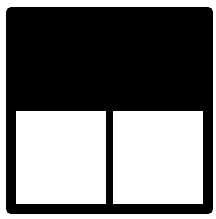}}}}
\newcommand{\fatcheckbv}{{\raisebox{-.3em}{\includegraphics[height=1.2em]{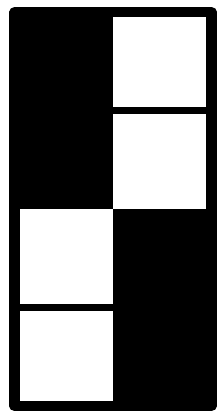}}}}
\newcommand{\variant}{{\opmath{\Phi}}}
\newcommand{\pot}{{v}}

\newcommand{\hc}{{\ensuremath{\hat c}}}
\newcommand{\bbT}{{\ensuremath{\mathbb{T}}}}

\newcommand{\Dvar}[1]{{\ensuremath{\operatorname{\Delta\!\variant_{#1}}}}}

\newcommand{\ie}{\textit{i.e.}\/}

\newcommand{\resp}{{resp.}\/}

\renewcommand{\paragraph}[1]{
	\noindent\textbf{#1}\quad}

\title{Progresses in the Analysis of\\ Stochastic 2D Cellular Automata: \\ a Study of Asynchronous 2D Minority
}
\author{Damien Regnault$^{1,2}$ 
\and Nicolas Schabanel$^{1,2}$
\and Éric Thierry$^{1}$}
\date{\small $^1$ \href{http://www.ixxi.fr/}{IXXI}-\href{http://www.ens-lyon.fr/LIP/index.php.en}{LIP}, École Normale Supérieure de Lyon, 46 allée d'Italie, 69364 Lyon Cedex 07, France. \href{http://perso.ens-lyon.fr/damien.regnault/}{\texttt{http://perso.ens-lyon.fr/\{damien.regnault}},\href{http://perso.ens-lyon.fr/eric.thierry/}{\texttt{eric.thierry\}}}. \\[1mm]
$^2$ \href{http://www.cmm.uchile.cl/}{CNRS, Centro de Modelamiento Matemático}, Universidad de Chile, Blanco Encalada 2120 Piso 7, Santiago de Chile. \href{http://www.cmm.uchile.cl/~schabanel}{\texttt{http://www.cmm.uchile.cl/$\sim$schabanel}}.}

\newtheorem{theorem}{Theorem}
\newtheorem{lemma}[theorem]{Lemma}
\newtheorem{corollary}[theorem]{Corollary}

\newtheorem{proposition}[theorem]{Proposition}

\newtheorem{definition}{Definition}
\newtheorem{example}{Example}

\newenvironment{Proof}{%
	\paragraph{\mdseries\itshape Proof.}%
	}{$\Box$\smallskip}

\begin{document}

\maketitle


\begin{abstract}
Cellular automata are often used to model systems in physics, social sciences, biology that are inherently asynchronous. Over the past 20 years, studies have demonstrated that the behavior of cellular automata drastically changed under asynchronous updates. Still, the few mathematical analyses of asynchronism focus on one-dimensional probabilistic cellular automata, either on single examples  or on specific classes. As for other classic dynamical systems in physics, extending known methods from one- to two-dimensional systems is a long lasting challenging problem. 

In this paper, we address the problem of analysing an apparently simple 2D asynchronous cellular automaton: 2D \Minority\/ where each cell, when fired, updates to the minority state of its neighborhood. Our experiments reveal that in spite of its simplicity, the minority rule exhibits a quite complex response to asynchronism. By focusing on the fully asynchronous regime, we are however able to describe completely the asymptotic behavior of this dynamics as long as the initial configuration satisfies some natural constraints. Besides these technical results, we have strong reasons to believe that our techniques relying on defining an energy function from the transition table of the automaton may be extended to the wider class of threshold automata. 
\smallskip

\centerline{\itshape An abstract version of this paper has been published in \cite{RST2007-VN-MFCS}.}
\end{abstract}

\section{Introduction}
\label{sec:intro}

In the literature, cellular automata have been both studied as a model of computation presenting massive parallelism, and used to model phenomena in physics, social sciences, biology... Cellular automata have been mainly studied under synchronous dynamics (at each time step, all the cells update simultaneously). But real systems rarely fulfill this assumption and the cell updates rather occur in an asynchronous mode often described by stochastic processes. Over the past 20 years, many empirical studies~\cite{Ber94,Ingerson84,FM04,Lumer1994,Sch99} have been carried out showing that the behavior of a cellular automaton may widely vary when introducing asynchronism, thus strengthening the need for theoretical framework to understand the influence of asynchronism. Still, the few mathematical analyses of the effects of asynchronism focus on one-dimensional probabilistic cellular automata, either on single examples like~\cite{Fuks02,Fuks04,R2006} or on specific classes like~\cite{FMST-TCS2006,FRST2006}. As for other classic dynamical systems in physics, such as spin systems or lattice gas, extending known methods from one- to two-dimensional systems is a long lasting challenging problem. For example, understanding how a configuration all-up of spins within a down-oriented external field evolves to the stable configuration all-down has only recently been solved mathematically and only for the limit when the temperature goes to $0$, \ie, when only one transition can occur at time (see \cite{ModelIsing}). Similarly, the resolution of the study of one particular 2D automaton under a given asynchronism regime is already a challenge.

\smallskip

\paragraph{Our contribution.} 
In this paper, we address the problem of understanding the asynchronous behavior of an apparently simple 2D stochastic cellular automaton: 2D \Minority\/ where each cell, when fired, updates to the minority state of its neighborhood. We show experimentally in Section~\ref{sec:experiment} that in spite of its simplicity the minority rule exhibits a quite complex response to asynchronism. We are however able to show in Section~\ref{sec:analysis} that this dynamics almost surely converges to a stable configuration (listed in Proposition~\ref{prop:stable}) and that if the initial configuration satisfies some natural constraints, this convergence occurs in polynomial time (and thus is observable) when only one random cell is updated at a time. Our main result (Theorems~\ref{thm:convergence:even} and~\ref{thm:conv:bounded}) rely on extending the techniques based on one-dimensional random walks developed in \cite{FMST-TCS2006,FRST2006} to the study of the two-dimensional random walks followed by the boundaries of the main components of the configurations under asynchronous updates. We have strong reasons to believe that our techniques relying on defining an energy function from the transition table of the automaton may be extended to the wider class of threshold automata.

Our results are of particular interest for modeling regulation network in biology. Indeed, 2D \Minority\/ cellular automaton represents an extreme simplification of a biological model where the biological cells are organized as a 2D grid and where the regulation network involves only two genes (the two states) which tend to inhibit each other \cite{Aracena2003}. The goal is thus to understand how the concentrations of each gene evolve over time within the biological cells, and in particular, which gene ends up dominating the other in each cell, \ie, in which state ends up each cell. Understanding this simple rule is thus a key step in the understanding of more complex biological systems. 






\section{Experimental results}
\label{sec:experiment}

\begin{figure}[t]
\centerline{\includegraphics[width=\textwidth]{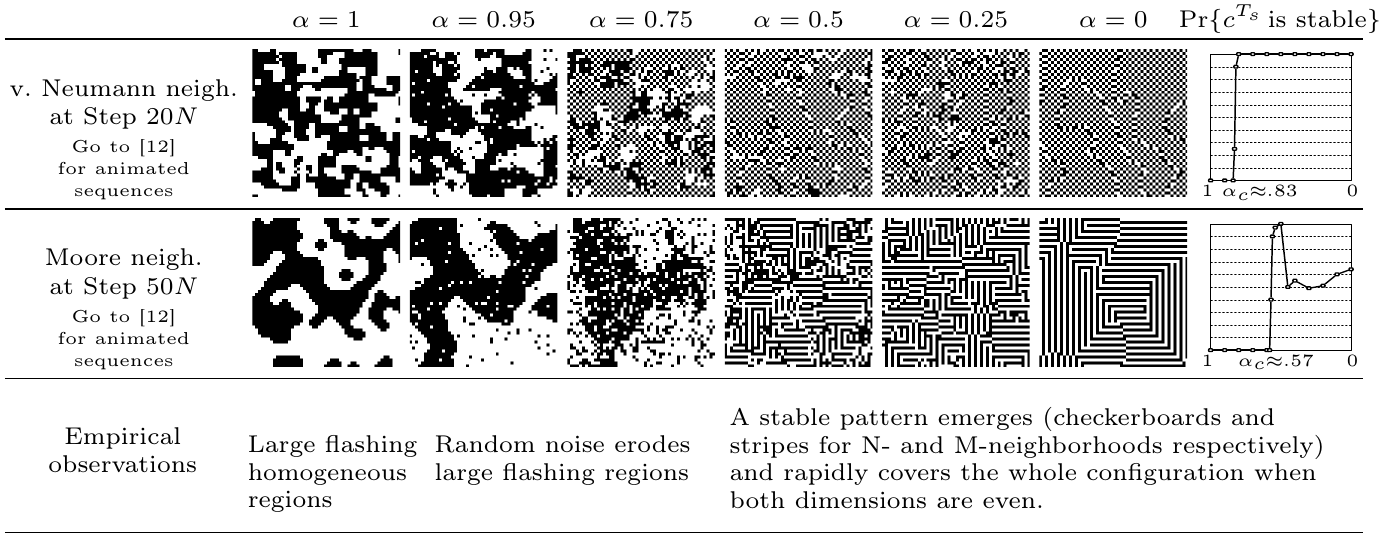}}
\caption{2D \Minority\/ under different $\alpha$-asynchronous dynamics with $N_{50}=50\times50$ cells. The last column gives, for $\alpha\in[0,1]$, the empirical probability that an initial random configuration converges to a stable configuration before time step $T_s\cdot N_{50}$ where $T_s = 1000$ and $T_s = 2000$ for von Neumann and Moore neighborhood respectively.}
\vspace*{-5mm}
\label{fig:experiments}
\end{figure}

This section is voluntarily informal because it presents experimental observations whose formalizations are already challenging open questions. The next section will present in a proper theoretical framework our progresses in the understanding of these phenomena. The configurations studied here consist in a set of cells organized as a $n\times m$ torus ($n$ and $m$ are even) in which each cell can take two possible states: $0$ (white) or $1$ (black). The asynchronous behavior of 2D minority automaton turns out to be surprisingly complex for both of the studied neighborhoods:
\begin{itemize}
\item \emph{von Neumann} (N-neighborhood for short), where each selected cell  updates to the minority state within itself and its neighbors N, S, E, and W; and
\item \emph{Moore} (M-neighborhood for short), where each selected cell updates to the minority state among itself and its 8 closest neighbors N, S, E, W, NE, NW, SE, and SW. 
\end{itemize}
In this section, we present a report on extensive experiments conducted on 2D Minority for both N- and M-neighborhood. 
 
In this section, we consider the \emph{$\alpha$-asynchronous 2D Minority} dynamics in which at each time step, each cell updates to the minority state in its own neighborhood independently with probability $\alpha$. We denote by $\alpha=0$ the \emph{fully asynchronous 2D Minority} dynamics in which at each time step, a daemon selects uniformly at random one cell and updates it to the minority state in its neighborhood. 

\paragraph{The synchronous regime}\!\!\!\!\!\!  ${(\alpha=1)}$ of 2D \Minority\/ has been thoroughly studied in \cite{GM90} where it is proved that it converges to cycles of length $1$ or $2$. Experimentally, from a random configuration, the synchronous dynamics in both neighborhoods converges to sets of large flashing  white or black regions.

\paragraph{As soon as a little bit of asynchronism}\!\!\!\!\!\! is introduced, the behavior changes drastically for both neighborhoods 
(see Fig.~\ref{fig:experiments} and open our website~\cite{WebsiteMovies} for animated sequences). 
Due to the asynchronism at each step, some random cells do not update and this creates a \emph{noise} that progressively erodes the flashing homogenous large regions that were stable in the synchronous regime. After few steps, the configuration seems to converge rapidly to a homogeneous flashing background perturbed by random noise.

\paragraph{Experiments provide evidences that there exists a threshold $\pmb{\alpha_c}$}\!\!\!\!\!\!, ${\alpha_c \approx .83}$ and ${\alpha_c \approx\! .57}$ for the N- and M-neighborhoods respectively, such that if $\alpha\leq\alpha_c$, then stable patterns arise (\emph{checkerboards} and \emph{stripes} for N- and M-neighborhood respectively). As it may be observed in 
\cite{WebsiteMovies}, above the threshold, when $\alpha > \alpha_c$, these patterns are unstable, but  below and possibly at $\alpha_c$, these patterns are sufficiently stable to extend and ultimately cover the whole configuration.

\paragraph{Convergence in asynchronous regimes.} The last column of Fig.~\ref{fig:experiments} shows that experimentally, when $\alpha\leq \alpha_c$, the asynchronous dynamics appears to converge at least with constant probability, rapidly to very particular stable configurations tiled by simple patterns known to be stable for the dynamics. Above the threshold, when $\alpha_c <\alpha < 1$, the asynchronous dynamics appears experimentally to be stuck into randomly evolving configurations in which no structure seems to emerge. 

We will show in Theorem~\ref{thm:convergence:even} that if at least one of the dimensions is even, the dynamics will almost surely reach a stable configuration, for all $0\leq \alpha < 1$, but after at most an exponential number of steps. We conjecture that below the threshold $\alpha_c$ this convergence occurs in polynomial time on expectation if both dimensions are even (the threshold $T_s=2000$ is probably too low for the M-neighborhood in Fig.\ref{fig:experiments}). We will prove this result in Theorem~\ref{thm:conv:bounded} for the fully asynchronous regime under the N-neighborhood under certain natural constraint on the initial configuration. Similar results to the ones to be presented below have been obtained in \cite{RST2007-Moore} for the M-neighborhood by extending of the techniques presented here.

\section{Analysis of fully asynchronous 2D Minority}
\label{sec:analysis}

 We consider now the fully asynchronous dynamics of 2D \Minority\/ with von Neumann neighborhood. Let $n$ and $m$ be two positive integers and $\bbT=\mathbb Z_n\times \mathbb Z_m$ the $n\times m$-torus. A \emph{$n\times m$-configuration} $c$ is a function $c:\bbT\rightarrow\{0,1\}$ that assigns to each \emph{cell} $(i,j)\in \bbT$ its \emph{state} $c_{ij}\in\{0, 1\}$ ($0$ is white and $1$ is black in the figures). We consider here the \emph{von Neumann neighborhood}: the \emph{neighbors} of each cell $(i,j)$ are the four cells $(i\pm1,j)$ and $(i,j\pm1)$ (indices are computed modulo $n$ and $m$, we thus consider periodic boundary conditions). We denote by $N=nm$, the total number of cells.


\begin{definition}[Stochastic 2D Minority]  
We consider the following dynamics $\dmin$ that associates to each configuration $c$ a random configuration $c'$ obtained as 
follows: a cell $(i,j)\in \bbT$ is selected uniformly at random and its state is updated to the minority state in its neighborhood (we say that cell $(i,j)$ is \emph{fired}), all the other cells remain in their current state:
$$
c'_{ij} = \left\{ \begin{array}{cl} 1 & \textup{if $c_{ij}+c_{i-1,j}+c_{i+1,j}+c_{i,j-1}+c_{i,j+1}\leq 2$}\\[1mm] 0 & \textup{otherwise}\end{array}\right.
$$ 
and $c'_{kl} = c_{kl}$ for all $(k,l)\neq(i,j)$. We say that a cell is \emph{active} if its neighborhood is such that  its state changes when the cell is fired. 
\end{definition}


\begin{definition}[Convergence]
We denote by $c^t$ the random variable for the configuration obtained from a configuration $c$ after $t$ steps of the dynamics: ${c^t = \dmin^t (c)}$; $c^0 = c$ is the \emph{initial configuration}.

We say that the \emph{dynamics $\dmin$ converges} almost surely from an initial configuration $c^0$ to a configuration $\bar c$ if the random variable $T=\min\{t:c^t = \bar c\}$ is finite with probability $1$. We say that the convergence occurs in polynomial (resp., linear, exponential) time on expectation if $\expect[T] \leq p(N)$ for some polynomial (resp., linear, exponential) function $p$.   
\end{definition}


\begin{figure}[t]

\includegraphics[width=\textwidth]{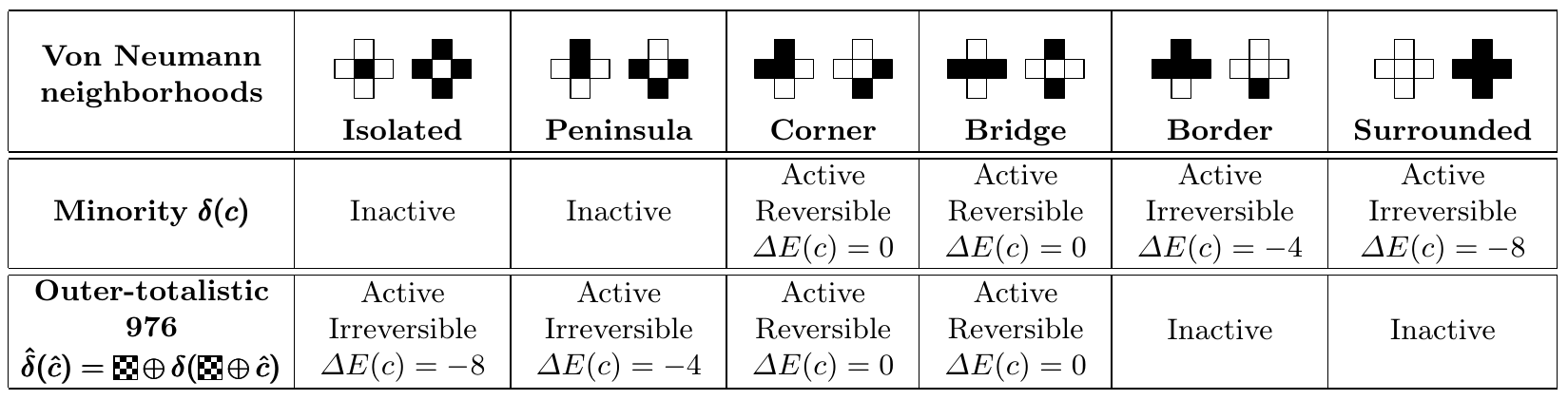}\\
\caption{Neighborhood's names and transition tables of \Minority\/ $\dmin$ and its counterpart \OT\/ $\dout$ (see section~\ref{sec:OT976}): only active cells switch their states when fired.}
\label{fig:neighborhood}
\end{figure}

As seen in Section~\ref{sec:experiment}, any configuration  tend to converge under this dynamics towards a \emph{stable configuration}, \ie, towards a configuration where all cells are in the minority state of their neighborhood, \ie, inactive. 
\smallskip

\paragraph{Checkerboard patterns.} 
We say that a subset of cells~$R\subseteq \bbT$ is \emph{connected} if $R$ is connected for the neighborhood relationship. 
We say that $R$ is \emph{checkerboard-tiled} if all adjacent cells in $R$ are in opposite states. 
A  \emph{horizontal} (\resp, \emph{vertical}) \emph{band} of width $w$ is a set of cells $R=\{(i,j): k\leq i < k+w\}$ for some $k$ (\resp, $R=\{(i,j): k\leq j < k+w\}$). 

\subsection{Energy of a configuration}

The following natural parameters measure the stability of a configuration, \ie, how far the cells of the configuration are from the minority state in their neighborhood. Following the seminal work of Tarjan in amortized analysis \cite{TarjanAmortizedPotential}, we define a local potential that measures the amount of  local unstability in the configuration. We proceed by analogy with the spin systems in statistical physics (Ising Model \cite{ModelIsing}): we assign to each cell a potential equal to the benefit of switching its state; this potential is naturally defined as the number of its adjacent cells to which it is opposed (\ie, here, the number of cells which are in the same state as itself); summing the potentials over all the cells defines the total energy of the configuration at that time. As we consider arbitrary initial configuration, the system evolves \emph{out-of-equilibrium} until it (possibly) reaches a stable configuration, thus its energy will vary over time; in particular, as will be seen in Proposition~\ref{prop:energy:decreases}, its energy will strictly decrease each time an irreversible transition is performed (\ie, each time a cell of potential $\geq 3$ is fired). It turns out that this energy function plays a central role in defining, in Section~\ref{sec:variant}, the variant that will be used to prove the convergence of the system. We will see in particular that as observed experimentally in Section~\ref{sec:experiment}, the system tends to reach configurations of minimal energy as one would expect in a real physical system.


\begin{definition}[Energy]
The \emph{potential} $\pot_{ij}$ of cell $(i,j)$ is the number of its four adjacent cells that are in the same state as itself. The \emph{energy} of a configuration $c$ is defined as the sum of the potentials of the cells: $E(c) = \sum_{i,j} v_{ij}$. 
\end{definition}

\begin{definition}[Borders]
We say that there is a \emph{border} between two neighboring cells if they are in the same state, \ie: 
\begin{itemize}
\item the edge between cells $(i,j)$ and $(i,j+1)$  is an horizontal border if $c_{ij} = c_{i,j+1}$;
\item the edge between cells $(i,j)$ and $(i+1,j)$ is a vertical border if $c_{ij} = c_{i+1,j}$.
\end{itemize}

\end{definition}

\begin{definition}[Homogeneous regions]
An \emph{alternating path} is a sequence of neighboring cells that does not go through a border, \ie, of alternating states. This defines an equivalence relationship «\,being connected by an alternating path\,», the equivalence classes of this relationship are called the \emph{homogenous regions} of the configuration. 
\end{definition}

\begin{proposition}
Each homogeneous region is connected and tiled by one of the two checkerboard patterns, either \checkb\/ or \checkbb. The boundary of each homogeneous region is exactly the set of borders touching its cells.  
\end{proposition}

\begin{proposition}
The potential of a cell is the number of borders among its sides. The energy of a configuration is twice the number of borders. 

A cell is active if and only if at least two of its sides are borders.
\end{proposition}

\begin{corollary}
If both dimensions $n$ and $m$ have the same parity, $(\forall c)\, E(c)\in 4\mathbb N$; and $(\forall c)\, E(c)\in 2+4\mathbb N$ otherwise. 
\end{corollary}

The energy of a $n\times m$-configuration belongs to $\{0,2,4,\ldots,4N\}$ since each pair of adjacent cells in the same state are counted twice and ${0\leq v_{ij}\leq 4}$ for all $(i,j)$.
There are two configurations of maximum energy $4N$: \emph{all-black} and \emph{all-white}. If $n$ and $m$ are even, there are two configurations of energy zero: the two \emph{checkerboards}. If $n$ is even and $m$ is odd, the minimum energy of a configuration is $2n$ and such a configuration consists in a checkerboard pattern wrapped around the odd dimension creating a vertical band of width $2$ tiled with pattern \linecheckb.

\smallskip

\paragraph{Energy of stable configurations.} A cell is inactive if and only if its potential is $\leq 1$. It follows that the energy of any stable configuration belongs to $\{0,2,\ldots,N\}$. Stable configurations are thus as expected of lower energy. If $n$ and $m$ are even and at least one of them is a multiple of $4$, there are stable configurations of maximum energy $N$, tiled by the ``fat''-checkerboard \fatcheckb\/ or \fatcheckbv. 

\smallskip

\paragraph{Energy is non-increasing.} Under the fully asynchronous dynamics $\dmin$, the energy may not increase over time.


\begin{proposition}
\label{prop:energy:decreases}
From any initial configuration $c$, the random variables $E(c^t)$ form a non-increasing sequence and $E(c^t)$ decreases by at least $4$ each time a cell of potential $\geq 3$ is fired. 
\end{proposition}

\begin{Proof}
The variation of the energy of the configuration when the state of a cell of potential $v$ is flipped is $8-4v\leq 0$, since active cells have potential $\geq 2$.
\end{Proof}


\paragraph{Initial energy drop.} Furthermore, after a polynomial number of steps and from any \emph{arbitrary} initial configuration, the energy falls rapidly below $5N/3$, which is observed experimentally through the rapid emergence of checkerboard patterns in the very first steps of the evolution: 

\begin{proposition}[Initial energy drop]
\label{prop:energy:drop}
The random variable ${T = \min\{t:E(c^t) < 5N/3\}}$ is almost surely finite and $\expect[T] = O(N^2)$. 
     
\end{proposition}

\begin{Proof}
Consider a configuration $c$ with energy $E>5N/3$. We will show that either $c$ contains a cell of potential $\geq 3$ or two adjacent cells of potential $2$ in opposite states. Let us proceed by contradiction and assume that every cell of $c$ has potential $\leq 2$ and that every adjacent cells of potential $2$ are in the same state. Let $b_{1^-}$, $b_2$ (resp. $w_{1^-}$ and $w_2$) be the number of black (resp. white) cells of potential $\leq 1$ and $2$. Let consider the bipartite graph that connects each black cell of potential $2$ to its adjacent white cells of potential $\leq 1$. Every black cell of potential $2$ is adjacent to exactly $2$ white cells of potential $\leq 1$ and every white cell of potential $\leq 1$ is adjacent to at most $4$ black cells of potential $2$. The number of edges in the bipartite graph is thus at least $2b_2$ and at most $4w_{1^-}$, it follows that $2b_2\leq 4 w_{1^-}$. Symmetrically, $2w_2\leq 4b_{1^-}$. But, $N = b_{1^-}+b_2+w_{1^-}+w_2 \geq 3(b_2+w_2)/2$, thus the configuration admits at most $2N/3$ cells of potential $2$ and its energy is  $\leq N+2N/3$, contradiction. 

Consider now the variant $\Psi(c^t)=3E(c^t)/2-N_{3^+}(c^t)$ where $N_{3^+}(c^t)$ is the number of cells of potential~$\geq 3$ in $c^t$. For all time $t$, $0 \leq \Psi(c^t) \leq 6N$. Let $N_{2}$ be the number of pairs of adjacent cells with potential~2 in opposite states. $E(c^t)$ is a non-decreasing function of time and each time a cell of potential $\geq 3$ is fired, $E(c^t)$ decreases by at least $4$; it follows that $\expect[E(c^{t+1})-E(c^t)] \leq -4N_{3^+}(c^t)/N$. A cell of potential $3$ may disappear only if itself or one of its four neighbors are fired; and each time a cell of potential $2$ adjacent to a cell of potential $2$ in an opposite state is fired, the potential of the later cell increases to $3$. It follows that:
$$
\expect[N_{3^+}(c^{t+1})-N_{3^+}(c^t)]\geq \frac{N_2(c^t)-5N_{3^+}(c^t)}{N}.
$$
Summing up the two terms yields:
$$
\expect[\Psi(c^{t+1})-\Psi(c^t)]\leq -\frac{N_{3^+}(c^t)+N_{2}(c^t)}N.
$$
Then, as long as $E(c^t)\geq 5N/3$, $\Psi(c^t)$ decreases at each time step by at least $1/N$ on expectation. Since $\Psi$ is bounded by $6N$, a classic stopping time analysis  (see for example, Lemma~2 in~\cite{FMST-TCS2006}) shows that after at most $O(N^2)$ steps on expectation, either $E(c^t)$ drops below~$5N/3$ or $\Psi(c^t)$ drops below~$3N/2$ which also implies that $E(c^t)\leq 5N/3$. 
\end{Proof}

%


\subsection{Stable configurations}

\begin{proposition}[Stable configurations]
\label{prop:stable}
Stable configurations are the configurations composed of checkerboard-tiled bands. More precisely: 
\begin{itemize}
\item 
\emph{if $n$ or $m$ is even,} the stable configurations are the configurations composed of a juxtaposition of horizontal bands (or of vertical bands) of width $\geq 2$ tiled by checkerboards; 
\item
\emph{if $n$ (\resp, $m$) is odd and $m$ (\resp, $n$) is even,} the bands are necessarily horizontal (\resp, vertical);
\item
\emph{finally, if $n$ and $m$ are odd,} no stable configuration exists.
\end{itemize}
\end{proposition}

\begin{Proof}
In a stable configuration, every cell touches at most one border. It follows that borders of the homogeneous regions form straight lines at least 2 cells apart from each other.
\end{Proof}

\begin{figure}
\centerline{\begin{tabular}{c @{~~~~~~~~~~~~} c} 
\begin{tabular}[t]{cccc}
\includegraphics[height=2cm]{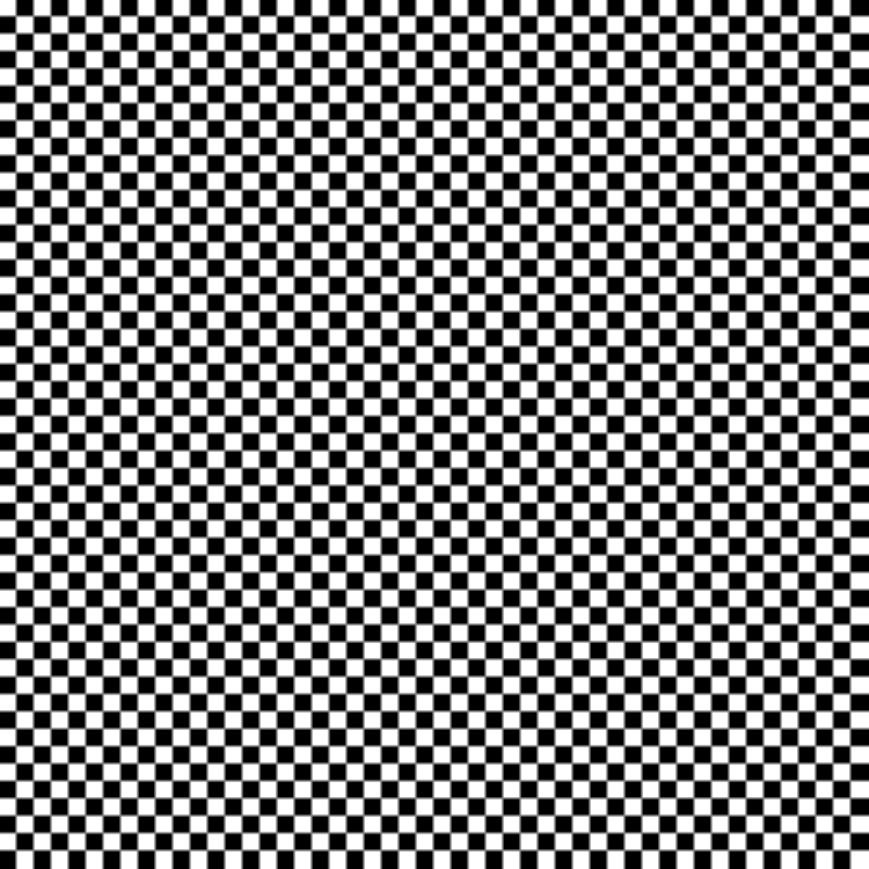}
&\includegraphics[height=2cm]{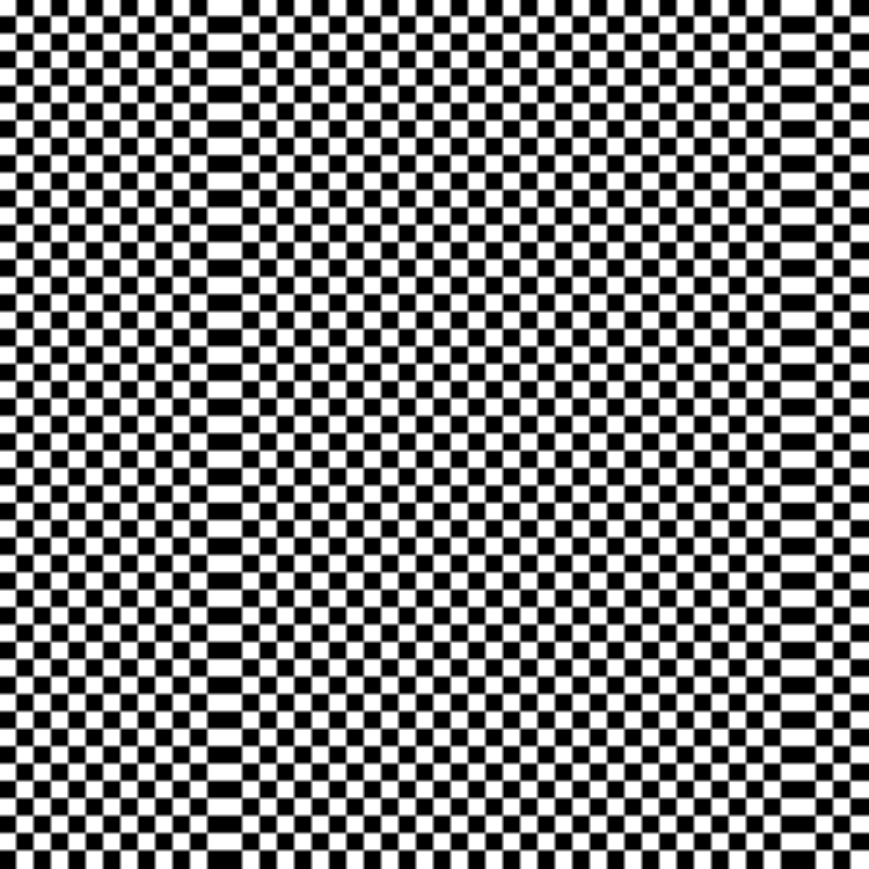}
&\includegraphics[height=2cm]{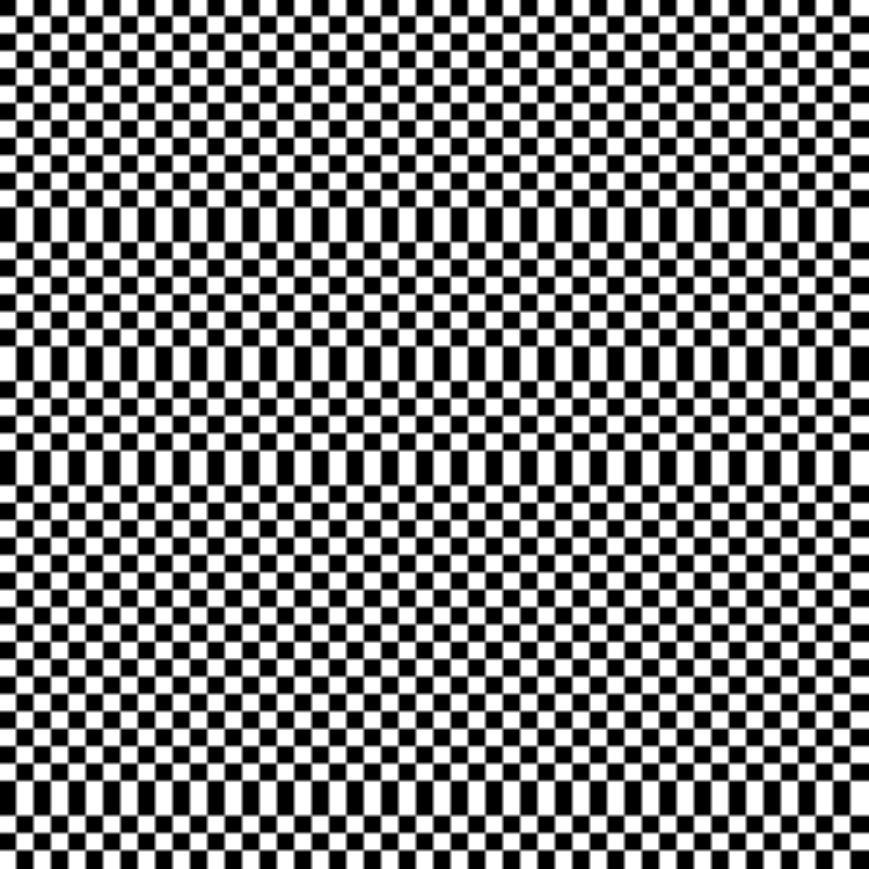}
&\includegraphics[height=2cm]{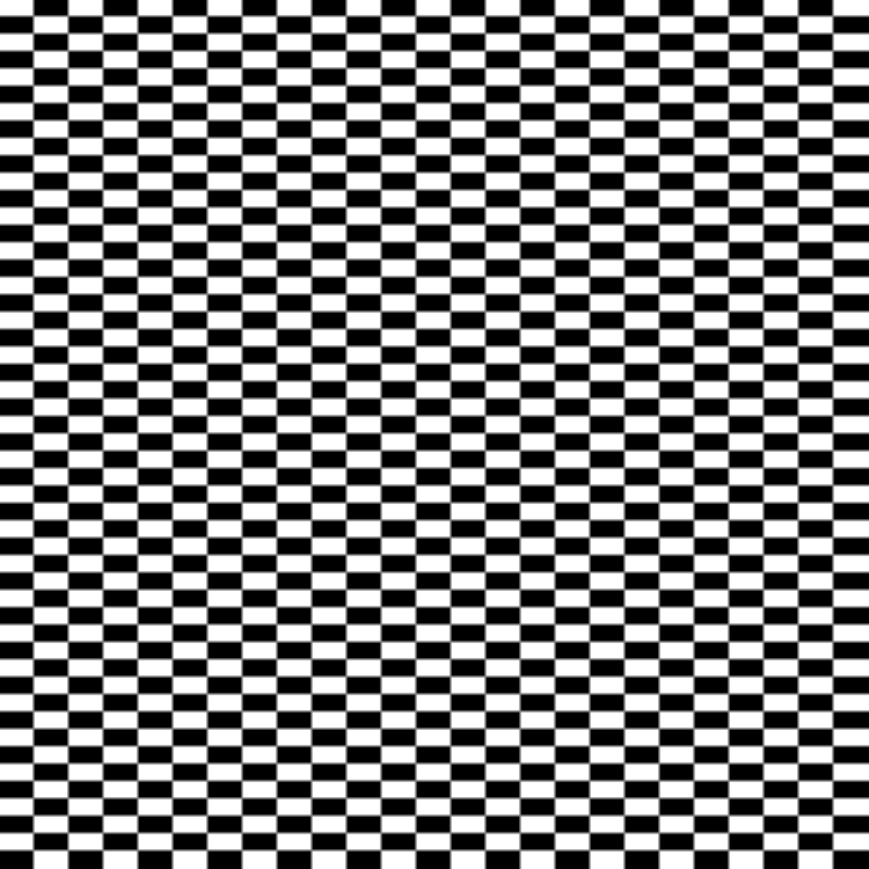}\\
$E = 0$
&
&
&
$E = 4N$
\end{tabular}
&
\begin{tabular}[t]{cc}
\includegraphics[height=2cm]{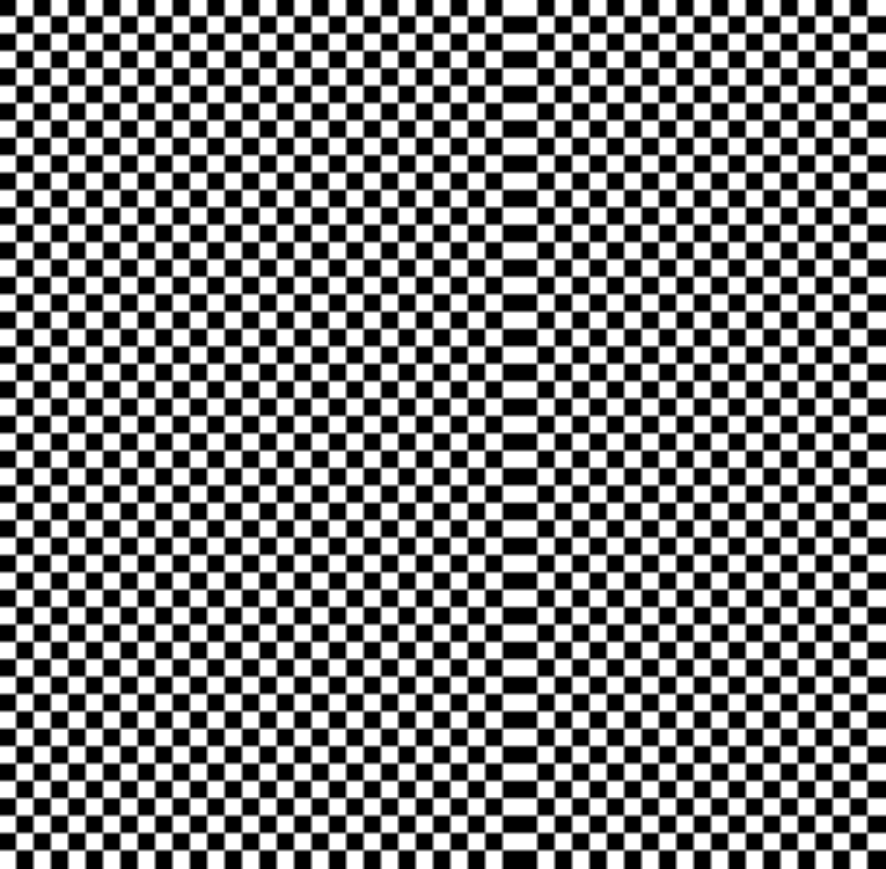}
&\includegraphics[height=2cm]{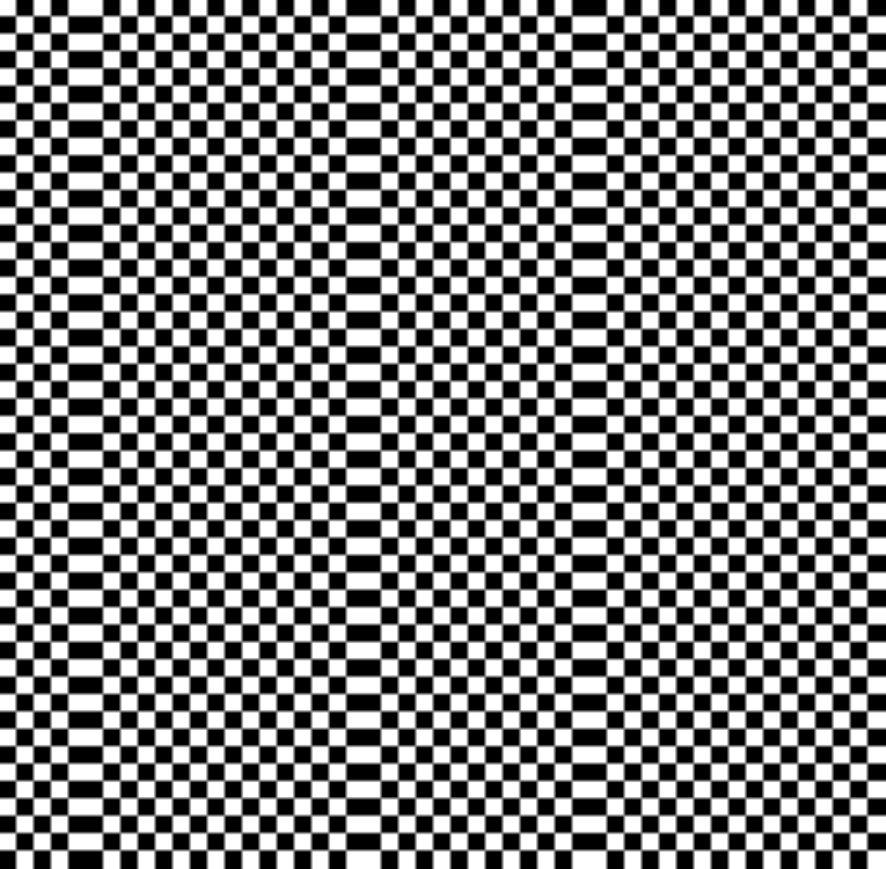}
\end{tabular}
\\
a) $n$ and $m$ are even & b) only $n$ is even\\
\end{tabular}
}
\caption{Examples of stable configurations.}
\label{fig:stable}
\end{figure}

\begin{corollary}
If $n$ and $m$ are odd, the dynamics $\delta$ never reaches a stable configuration.
\end{corollary}


\subsection{Coupling with \OT}
\label{sec:OT976}

From now on up to the end of section~\ref{sec:analysis}, we assume that $n$ and $m$ are even (with the only exception of Corollary~\ref{cor:convergence:odd}). We denote by \checkb\/ the checkerboard configuration of energy $0$ defined as follows: $\checkb_{\,ij} = (i+j)\mod2$. Given two configurations $c$ and $c'$, we denote by $c\XOR c'$ the \textsc{xor} configuration $c''$ such that $c''_{ij} = (c_{ij} + c'_{ij}) \mod 2$.

\smallskip

\paragraph{Dual configurations.}  As observed above, the fully asynchronous dynamics $c^t$ tends to converge from any initial configuration $c^0$ to configurations tiled by large checkerboard regions. It is thus convenient to consider instead, the sequence of \emph{dual configurations} $(\hc^t)$ defined by $\hc^t = \checkb \XOR c^t$, in which the large checkerboard regions of $c^t$ appear as large homogeneous black or white regions. Clearly, the dual sequence $\hc^t$ evolves according to the dynamics $\dout(.) = \checkb \XOR \dmin(\checkb \XOR .)$, indeed for all~$t$, $\hc^{t+1} = \checkb \XOR c^{t+1} = \checkb \XOR \dmin(c^t) = \checkb \XOR \dmin(\checkb \XOR \hc^t) = \dout(\hc^t).$\\ By construction, the two dual random sequences $(c^t)$ and $(\hc^t)$ as well as their corresponding dynamics $\dmin$ and $\dout$ are \emph{coupled probabilistically} (see \cite{Coupled}): the \emph{same} random cell is fired in both configurations at each time step. A simple calculation shows that the dual dynamics $\dout$ associates to each dual configuration $\hc$, a dual configuration $\hc'$ as follows: select uniformly at random a cell $(i,j)$ (the same cell $(i,j)$ as $\dmin$ fires on the primal configuration~$c$) and set:

$$
\hc'_{ij} =
\left\{ 
\begin{array}{c@{~~}l} 
1 & \textup{if $\Sigma \geq 3$}\\[1mm] 
1-\hc_{ij} & \textup{if $\Sigma = 2$}\\[1mm]
0 & \textup{otherwise}\end{array}\right.
\text{ with $\Sigma = \hc_{i-1,j}+\hc_{i+1,j}+\hc_{i,j-1}+\hc_{i,j+1}$}
$$

and $\hc'_{kl} = \hc_{kl}$ for all $(k,l)\neq (i,j)$. It turns out that this rule corresponds to the asynchronous dynamics of the cellular automaton \OT\/ \cite{Wolfram-website}. The corresponding transitions are given in Fig.~\ref{fig:neighborhood}.

\paragraph{Stable configurations of \OT.} We define the energy of the dual configuration $\hc$ and the potentials of each of its cells $(i,j)$ as the corresponding quantities, $E(c)$ and $v_{ij}$, in the primal configuration~$c$. By Proposition~\ref{prop:stable},
%
the stable dual configurations under the dual dynamics $\dout$ are the dual configurations composed of homogeneous black or white bands of widths~$\geq 2$. The two dual configurations of minimum energy $0$ are all-white and all-black.   

\begin{figure}[t]
\includegraphics[width=\textwidth]{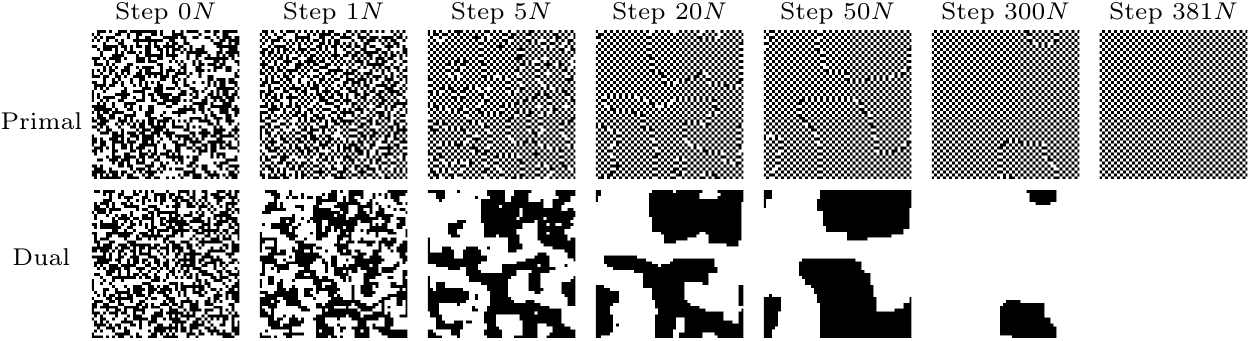}
\caption{The coupled evolutions of \Minority\/ \dmin\/ on the primal configurations $(c^t)$ (above) and its counterparts \OT\/ \dout\/ on dual configurations $(\hc^t)$ (below). Note that from step $50N$ on, $(c^t)$ an $(\hc^t)$ are bounded configurations.}
\vspace*{-5mm}
\label{fig:dmin:dout}
\end{figure}

Experimentally, any dual configuration under the fully asynchronous dynamics $\dout$ evolves towards large homogeneous black or white regions (corresponding to the checkerboard patterns in the primal configuration). Informally, these regions evolve as follows (see Fig.~\ref{fig:neighborhood}): isolated points tend to disappear as well as peninsulas; borders and surrounded points are stable; large regions are eroded in a random manner from the corners or bridges that can be flipped reversibly and their boundaries follow some kind of 2D random walks until large bands without corners ultimately survive (see Fig.~\ref{fig:dmin:dout} or 
\cite{WebsiteMovies}).

\subsection{Convergence from an arbitrary initial configuration}

In this section, we consider \emph{arbitrary} initial configurations $c^0$ and show that indeed the dynamics $\dmin$ converges to a stable configuration almost surely and after at most an exponential number of steps on expectation. 


\begin{theorem}
\label{thm:convergence:even}
From any initial configuration $c^0$, the dynamics $\dmin$ convergences to a stable configuration after at most $2N^{2N+1}$ steps on expectation. 
\end{theorem}

\begin{Proof}
According to the coupling above, it is equivalent to prove this statement for the dual dynamics. The following sequence of $\dout$-updates transforms any dual configuration~$\hc$ into a dual stable configuration : 

\begin{itemize}
\item \textbf{Phase I :} 
as long as there are active white cells, choose one of them and switch its state to black; 
\item \textbf{Phase II :}
as long as there are active black cells, choose one of them and switch its state to white.
\end{itemize}

During phase 1, the black regions expand until they fill their surrounding bands or surrounding rectangles. Clearly according to the transition table Fig.~\ref{fig:neighborhood}, after phase 1 of the algorithm, every white cell is inactive and thus is either a border or surrounded. In particular, no white band of width 1 survived. During phase 2, the black cells enclosed in rectangles or in bands of width 1 are eroded progressively and ultimately disapear. Finally, only black bands of width $\geq 2$ survive at the end of phase 2 and the configuration is stable since it is composed of homogeneous white or black bands of width $\geq 2$ (see Proposition~\ref{prop:stable}). During each phase, at most $N$ cells change their state. We conclude that, from any configuration $\hc$, there exists a path of length at most $2N$ to a stable configuration. Now, splits the sequence $(c^t)$ into segments $(c^{2Nk+1},...,c^{2N(k+1)})$ of length $2N$. The sequence of updates in each of these segments has a probability $1/N^{2N}$ to be the sequence of at most $2N$ updates given above that tranforms configuration $c^{2Nk}$ into a stable configuration. Since these events are independent, this occurs after $N^{2N}$ trials on expectation. We conclude that the dynamics $\dout$ and thus $\dmin$ converge to a stable configuration after at most $2N\cdot N^{2N}$ steps on expectation.%
\end{Proof}


\begin{corollary}
\label{cor:convergence:odd}
From any initial $n\times m$-configuration $c^0$, where $n$ is even and $m$ is odd, the dynamics $\dmin$ convergences to a stable configuration after at most $3N^{3N+1}$ steps on expectation. 

\end{corollary}

\begin{Proof}
Consider the cells within the $n\times (m-1)$ rectangle excluding the last column $m-1$. Consider the dual configuration inside this rectangle and apply the same sequence of updates as above. After Phase I, the black regions within the rectangle have been extended up to their bounding rectangles and furthermore no proper white horizontal band remains because since $m$ is odd, either one of the white cells at the extremity of such a band would be active (whatever the states of the cells in the last column are). After Phase II, the black rectangles have been erased as well as the proper horizontal black bands (since $m$ is odd, either one of the cells at the extremities of such a band would be active). At this stage, the only remaining active cells are within the last column $m-1$ and possibly in either one of the two neighboring columns $0$ or $m-2$. An extra series of at most $2n$ updates allows then to stabilize the cells in these two columns. It follows that a sequence of at most $2N+2n\leq 3N$ updates stabilize any configuration, which concludes the result by the same argument as above.
\end{Proof}


\begin{example}[Conjecture]
\label{ex:conv:exp}
Draw a rectilinear gray line wrapped twice around the short odd dimension of a $(2n+1)\times 2n^3$-configuration. Cut the configuration along this line and tile the unwraped configuration with a checkerboard pattern. Once rewrapped, the only active cells of the configuration are along the gray line (see Fig.~\ref{fig:wrapped}).

\begin{figure}[htb]
\centerline{\includegraphics[width=.6\textwidth]{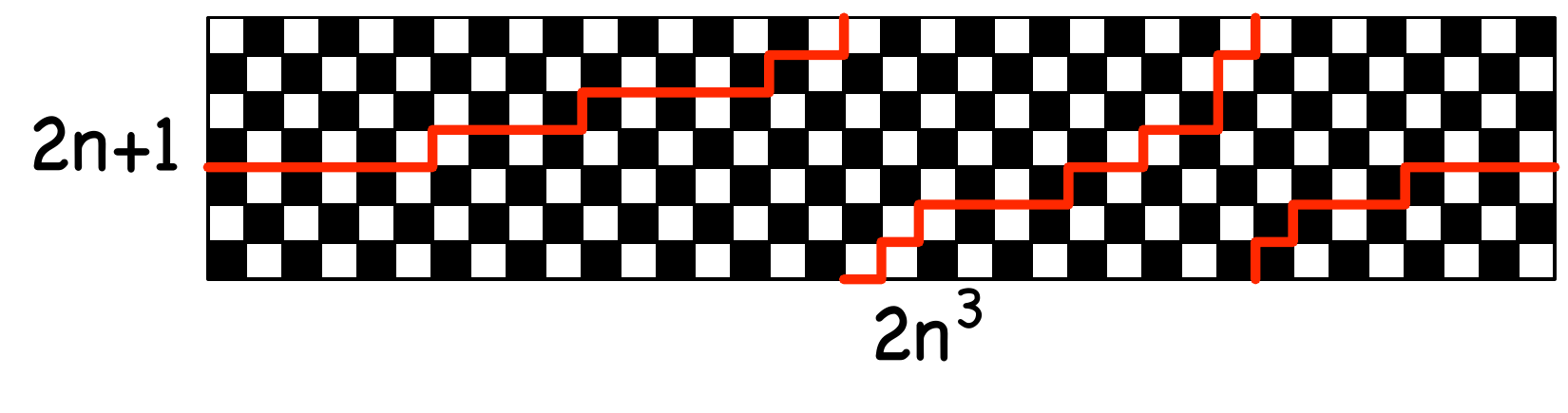}}\vspace*{-1em}
\caption{An odd\,$\times$\,even configuration with an exponential convergence time?} \label{fig:wrapped}
\end{figure}

The dynamics $\dmin$ can converge only after that  the gray line is unwrapped, \ie, only after it merges with itself somewhere, which is only possible if the line bends itself into a rectangle whose opposite corners meet at the same point of the torus. Unfortunately, the ``tension'' imposed by the wrapping around tends to spread apart the two parts of the gray line around the torus (in order to bend itself into a rectangle, the $n$ random walks of the corners on the gray line have to synchronize). We thus conjecture that this necessary self-crossing of the gray line may only occur after an exponential number of steps (which is confirmed by experiments). 
\end{example}


\subsection{Convergence from a bounded configuration}

We consider again that $n$ and $m$ are even. We observe experimentally that most of the time, the dynamics converges rapidly to one of the two checkerboard configurations of energy zero. We demonstrate in this section that if the dynamics reaches a configuration composed of an arbitrary region surrounded by a checkerboard, then it will converge to the corresponding checkerboard configuration almost surely after a polynomial number of steps on expectation. This corresponds to the analysis of the last steps of the behavior observed in experimentation. We believe that the techniques developed here may be extended to prove that the dynamics converges to a stable configuration in polynomial expected time from any initial configuration (see discussions in section~\ref{sec:conclusion}).


\begin{definition}[Bounded configuration]
We say that a configuration $c$ is \emph{bounded} if there exists a $(n-2)\times(m-2)$ rectangle such that the states in $c$ of the cells outside this rectangle are equal to the corresponding states in one of the two checkerboard configurations. W.l.o.g., we assume that the upper-left corner of the rectangle is $(1,1)$ and that the checkerboard is $\checkb$, \ie, a configuration $c$ is bounded if $c_{ij} = (i+j)\mod 2$ for all $(i,j)\in\{(i,j): (-1\leq i\leq 0) \textup{ or } (-1\leq j\leq 0)\}$.
\end{definition}


\begin{lemma}
\label{lem:dmin(convex)=convex}
If $c$ is a bounded configuration, $\dmin(c)$ is also bounded. 
\end{lemma}

\begin{Proof}
The cells belonging to the checkerboard pattern outside the rectangle have 3 adjacent cells in the state opposite to their own states; these cells are thus inactive (whatever the state of their adjacent cell inside the rectangle~is).
\end{Proof}


A bounded configuration is thus equivalent to a finite perturbation of an infinite planar configuration in $\mathbb Z^2$ tiled with the $\checkb$ pattern. Since the dual of $\checkb$ is the configuration all-white, the dual of a bounded configuration is thus equivalent to a \emph{finite number of black cells}, included into a $(n-2)\times(m-2)$ rectangle within an \emph{infinite white planar configuration in $\mathbb Z^2$}. We shall now consider this setting.


\begin{definition}[Convexity]
We say that a set of cells $R\subseteq \mathbb Z^2$ is \emph{convex} if for any pair of cells $(i,j)$ and $(i+k,j)$ (\resp, $(i,j+k)$) in $R$, the cells $(i+\ell,j)$ (\resp, $(i,j+\ell)$) for $0\leq \ell\leq k$ belong to $R$. We say that $R$ is an \emph{island} if $R$ is connected and convex.
\end{definition}


Our proof of the convergence of the dynamics in polynomial time for bounded configurations relies on the definition of a variant which decreases on expectation over time. It turns out that in order to define the variant, we do not need to consider the exact internal structure of the bounded configuration, but only the structure of the convex hull of its black cells. 


\begin{definition}[Convex hull of a configuration]
For any finite set of cells $R\in\mathbb Z^2$, we denote by $\hull(R)$ the convex hull of the cells in $R$, \ie, $\hull(R) = \cap\bigl\{S\subseteq \mathbb Z^2:\text{$S$ is convex and $S\supseteq R$}\}$. Given a bounded dual configuration $\hc$, we define the \emph{convex hull of $\hc$},  $\hull(\hc)$, as the dual configuration whose black cells are the cells in the \emph{convex hull} of the black cells of $\hc$, \ie, if $R = \{(i,j):\hc_{ij}=1\}$, $\hull(\hc)_{ij} = 1$ if and only if $(i,j)\in\hull(R)$. We say that a configuration $c$ is \emph{convex} if $\hc = \hull(\hc)$. 

We say that $\hc\leq \hc'$ if for all $(i,j)$, $c_{ij}\leq c'_{ij}$. Let $\hc$ be a convex dual bounded configuration. We define for each black cell $(i,j)$ in $\hc$, the \emph{island of $\hc$ that contains cell $(i,j)$}, as the maximum connected and convex configuration $\hc'$ such that $\hc'_{ij}=1$ and $\hc'\leq\hc$. This defines a unique \emph{decomposition into black islands} of the convex bounded configuration $\hc$.
\end{definition}


\paragraph{The variant.} \label{sec:variant}
We now consider the following \emph{variant}: $\variant(\hc) = E(\hull(\hc))/4 + |\hull(\hc))|$, where |\hull(\hc))| is the number of black cells in the convex hull configuration $\hull(\hc)$. We will show that from any initial configuration $c^0$, $\variant(c^t)$ decreases by at least $1/N$ on expectation at each time step until it reaches the value $0$, \ie, until the primal and dual configurations $c^t$ and $\hc^t$ converge to the infinite checkerboard and the infinite all-white configurations respectively. In order to prove that $\variant(c^t)$ decreases on expectation, we need to study the evolution of the convex hull of $\hc^t$; for this purpose, we introduce a modified coupled dual dynamics $\dhull$ that preserves the convexity of a dual configuration. Given a dual configuration $\hc$, we denote by $\dhull(\hc)$ the random configuration $\hc'$ such that: $\hc' = \dout(\hc)$ if the cell
 updated by $\dout$ is \emph{not} a black bridge, and $\hc' = \hc$ otherwise.


\begin{lemma}
\label{lem:dhull(convexebounded)}
If $\hc$ is a convex bounded configuration, $\dhull(\hc)$ is a convex bounded configuration.
\end{lemma}

\begin{Proof}
The only active transition in $\dout$ that would break the convexity of the black cells is updating a black bridge (see Fig.~\ref{fig:neighborhood}), but this transition is not allowed in $\dhull$.
\end{Proof}




\begin{lemma}
\label{lem:E:convex}
For all convex bounded configurations $\hc$ and $\hc'$, if $\hc \leq \hc'$, then $E(c)\leq E(c')$.
\end{lemma}

\begin{Proof}
The energy of a configuration $\hc$ is by definition twice the number of adjacent cells in opposite states in $\hc$, that is to say twice the number of sides of cells on the boundaries of the black islands that compose $\hc$, \ie, twice the sum of their perimeters. Since $\hc\leq\hc'$, the black islands that compose $\hc$ are included within the black islands that compose $\hc'$. Moreover, since the sets of rows and columns touched by the black islands that compose a convex configuration are pairwise disjoint, the sum of the perimeters of the black islands of $\hc$ that are included in the same black island of $\hc'$ is bounded from above by the perimeter of this later island.
\end{Proof}


The following lemma proves that the image of the convex hull of $\hc$ by the dynamics $\dhull$ bounds from above the convex hull of the image of $\hc$ by the dynamics $\dout$.


\begin{lemma}
\label{lem:dout<=dhull}
For all bounded configuration $\hc$, $\dout(\hc) \leq \dhull(\hull(\hc))$.
\end{lemma}

\begin{Proof}
We only need to prove that 1) if $\dout$ updates a white active cell in $\hc$, the corresponding cell in $\dhull(\hull(\hc))$ is black and 2) if $\dhull$ updates an active black cell in $\hull(\hc)$, then the corresponding cell in $\dout(\hc)$ is white. This is a direct consequence of the coupling of the dynamics of $\dout$ and $\dhull$.    

If a white active cell in $\hc$ is fired and if the corresponding cell in $(\hull(\hc))$ is white then both cells become black. If a white active cell in $\hc$ is fired and if the corresponding cell in $(\hull(\hc))$ is black then since the cell in $\hc$ is active it has two black neighbors thus the cell in $(\hull(\hc))$ has two black neighbors. The only kind of active cell with at least two neighbors of the same color under $\dhull$ dynamics is the corner cell. Indeed if a corner white cell in $\hc$ is black in $(\hull(\hc))$ then it is a border or surrounded cell. Thus if $\dout$ updates a white active cell in $\hc$, the corresponding cell in $\dhull(\hull(\hc))$ is black.

An active black cell in $\hull(\hc)$ under $\dhull$ dynamics is an active black cell in $\hc$ under $\dout$ dynamics. Thus if $\dhull$ updates an active black cell in $\hull(\hc)$, then the corresponding cell in $\dout(\hc)$ is white.
\end{Proof}


Let $\Dvar\lambda(\hc)$ be the random variable for the variation of the variant after one step of a dynamics $\lambda$ from a configuration $c$, \ie, $\Dvar\lambda(\hc) = \variant(\lambda(\hc))-\variant(\hc)$. 


\begin{corollary}
\label{cor:Dvardout<=Dvardhull}
For all bounded configuration $\hc$, $\Dvar\dout(\hc) \leq \Dvar\dhull(\hull(\hc))$.
\end{corollary}

\begin{Proof}
By definition, 
$$
\Dvar\dhull(\hull(\hc)) - \Dvar\dout(\hc) = \bigl(|\dhull(\hull(\hc))| - |\hull(\dout(\hc))|\bigr)+ \bigl(E(\dhull(\hull(\hc))) - E(\hull(\dout(\hc)))\bigr).
$$
According to lemma~\ref{lem:dout<=dhull}, 
$\hull(\dout(\hc))\leq \dhull(\hull(\hc))$ 
and thus 
$|\hull(\dout(\hc))| \leq |\dhull(\hull(\hc))|$. And by Lemma~\ref{lem:E:convex}, since both configurations are convex, $E(\hull(\dout(\hc))) \leq E(\dhull(\hull(\hc)))$.
\end{Proof}



\begin{lemma}
\label{lem:Dvar:island}
For all bounded configuration $\hc$ that consists of a unique black island, 
$$
-4/N\leq {\expect[\Delta\variant_\dhull(\hc)] \leq -3/N}.
$$
\end{lemma}

\begin{Proof}
Each active cell is fired with probability $1/N$. According to the dynamics of $\dhull$ (the same as the dynamics of $\dout$, Fig.~\ref{fig:neighborhood}, except that black bridges are inactive), if $\hc$ consists of an island of size $\geq 2$, 
\begin{align*}
\expect[\Delta\variant_\dhull(\hc)] 
    &\displaystyle = - \textstyle\frac1N\, \bigl({{\#\{\text{black corners}\}} + 2\,  {\#\{\text{black peninsulas}\}}\bigr) + \frac1N\,{\#\{\text{white corners}\}}} \\
    &\displaystyle  = -\textstyle\frac1N{{\#\{\text{salient angles}\}} + \frac1N{\#\{\text{reflex angles}\}}} = -\textstyle\frac{4}N,
\end{align*}

since ${\#\{\text{salient angles}\}} - {\#\{\text{reflex angles}\}} = 4$ for all convex rectilinear polygon. Finally, if $\hc$ consists of a unique (isolated) black cell, $\Dvar\dhull(\hc) = -3/N$.
\end{Proof}


\begin{lemma}
\label{lem:Dvar}
For any bounded not-all-white configuration $\hc$, $\expect[\Dvar\dout(\hc)]\leq -\ell/N$, where $\ell$ is the number of islands that compose $\hull(\hc)$. 
\end{lemma} 

\begin{Proof}
By Corollary~\ref{cor:Dvardout<=Dvardhull}, $\expect[\Dvar\dout(\hc)]\leq\expect[\Dvar\dhull(\hull(\hc))]$. By convexity of $\hull(\hc)$, the sets of rows and columns touched by the islands that compose $\hull(\hc)$ are pairwise disjoint. Thus, one can index the islands from 1 to $\ell$ from left to right, and the contacts between islands can only occur between two consecutive islands at the corners of their surrounding rectangles. Each contact creates at most two new active white cells that contribute for $+1/N$ each to $\expect[\Dvar\dhull(\hull(\hc))]$. The contribution of each island to $\expect[\Dvar\dhull(\hull(\hc))]$ is at most $-3/N$ according to Lemma~\ref{lem:Dvar:island}. It follows that $\expect[\Dvar\dhull(\hull(\hc))] \leq -3\ell/N+2(\ell-1)/N \leq -\ell/N$. 
\end{Proof}


\begin{theorem}
\label{thm:conv:bounded}
The fully asynchronous minority dynamics $\dmin$ converges almost surely from any initial bounded configuration $c$ to the stable configuration of minimum energy, $\checkb$, and the expected convergence time is $O(AN)$ where $A$ is the area of surrounding rectangle of the black cells in $\hc$. 
\end{theorem}

\begin{Proof}
Initially and for all time $t\geq 0$, $\variant(\hc^t) \leq 2 (n-2+m-2)+A\leq 2N+A$. As long as $\hc^t\not\equiv 0$, $\variant(\hc^t)>0$ and according to Lemma~\ref{lem:Dvar}, $\expect[\Dvar\dout(\hc^t)]\leq -1/N$. It follows that the random variable $T=\min\{t:\variant(\hc^t)\leq 0\}$ is almost surely finite and $\expect[T] = O(nA)$ (by applying for example Lemma~2 in \cite{FMST-TCS2006}); and at that time, $\hc^T$ and $c^T$ are the stable configurations all-white and~\checkb, respectively.                                  
\end{Proof}


\begin{example}[Worst case configurations] 
\label{ex:rectangle}
Consider the initial dual bounded $n\times n$-configuration $\hc$ consisting of a black $2\times(n-2)$ rectangle. The expected time needed to erase one complete line of the rectangle is at least $\Omega(nN) = \Omega(AN)$.
\end{example}

\begin{Proof}
Consider the initial dual bounded $n\times n$-configuration $\hc$ consisting of a black $2\times(n-2)$ rectangle. The first time the dynamics $\dout$ will erase a black cell in a given column, this black cell has to be a black corner, which was created by the erasure of one of its black neighbors in a adjacent column. The expected time between the erasures of the first black cells in a given column and of the first black cell in an adjacent columns is thus $\Omega(N)$ (the expected time to fire the new black corner) and the expected time needed to erase one complete line of the rectangle is at least $\Omega(nN) = \Omega(AN)$.
\end{Proof}



\section{Concluding remarks}
\label{sec:conclusion}

This paper proposes an extension to 2D cellular automata of the techniques based on random walks developped in \cite{FMST-TCS2006,FRST2006} to study 1D asynchronous elementary cellular automata. Our techniques apply as well with some important new ingredients, to the Moore neighborhood where the cell fired updates to the minority state within its height closest neighbors \cite{RST2007-Moore}. We believe that these techniques may extend to the wide class of threshold automata, which are of particular interest, in neural networks for instance. We are currently investigating refinements of the tools developed here, based on the  study of the boundaries between arbitrary checkerboard regions in order to try to prove that every \emph{arbitrary} $n\times m$-configuration converges to a stable configuration in a polynomial number of steps when $n$ and $m$ are both even (we conjecture a convergence in time $O(N^3)$ for non-bounded toric configurations of even dimensions). This result would conclude the study of this automaton under fully asynchronous dynamics. The experiments lead in Section~\ref{sec:experiment} exhibit an impressive richness of behavior for this yet apparently simple transition rule. An extension of our results to arbitrary $\alpha$-asynchronous regime is yet a challenging goal, especially if one considers that most of the results concerning spin systems or lattice gas (at the equilibrium) apply only to the limit when the temperature tends to $0$, \ie, when only one transition occurs at a time.  

\smallskip

{
\paragraph{Acknowledgements.} We would like to thank C. Moore, R. D'Souza and J. Crutchfield for their useful suggestions on the physics related aspects of our work.

\bibliographystyle{plain}
\bibliography{ICALP07}
}

\end{document}